\documentstyle[aps,prl,epsfig,amsfonts]{revtex}
\newtheorem{Lemma}{Lemma}
\newtheorem{Prop}{Proposition}
\newtheorem{Thrm}{Theorem}
\tighten
\draft
\begin{document}
\title{Exactly Solvable Single Lane Highway Traffic Model With Tollbooths}
\author{H.~F. Chau$^1$\footnote{Electronic address: hfchau@hkusua.hku.hk},
 Hao Xu$^{1,2}$ and L.-G. Liu$^2$}
\address{$^1$ Department of Physics, University of Hong Kong, Pokfulam Road,
 Hong Kong, China}
\address{$^2$ Department of Physics, Zhongshan University, Guangzhou,
 Guangdong 510275, China}
\date{\today}
\maketitle
\begin{abstract}
 Tolls are collected on many highways as a means of traffic control and
 revenue generation. However, the presence of tollbooths on highway surely
 slows down traffic flow. Here, we investigate how the presence of tollbooths
 affect the average car speed using a simple-minded single lane deterministic
 discrete traffic model. More importantly, the model is exactly solvable.
\end{abstract}
\medskip
\pacs{PACS numbers: 45.70.Vn, 05.60.-k, 89.40.+k}
 Modern computing power allows us to simulate highway and city traffic by
 looking at the microscopic behaviors of all cars. Perhaps the most well-known
 model of this kind is the cellular automaton based Biham-Middleton-Levine
 (BML) model of two-dimensional city traffic \cite{BML}. Numerical simulations
 of this model strongly suggest that a first order phase transition from the
 full-speed phase to the completely jamming phase occurs as the car density of
 the system increases. Moreover, this phase transition is primarily the result
 of the exclusion volume effect \cite{BML,Yung}. In spite of its simplicity,
 very little rigorous result is known for the BML model \cite{Ana}.
\par
 Different generations and variations of the BML model have been investigated
 in the literature \cite{Gen}. These generalizations focus on different aspects
 of the problem. They include the introduction of more realistic traffic rules
 \cite{Rules}, study of the effects of over-passes and faulty traffic lights
 \cite{Lights}, application of cellular automaton based traffic rules to single
 and multiple lane highway traffic \cite{Multi} and investigation of higher
 dimensional traffic behaviors \cite{High_D}. In particular, Nagel and
 Schreckenberg \cite{NS} proposed a model of (one-dimensional) highway traffic.
 In addition to the regular acceleration and deceleration, they model the
 realistic behavior of car drivers by allowing them to apply their breaks in a
 stochastic manner. Later on, Fukui and Ishibashi (FI) \cite{FI} investigated a
 simple-minded deterministic model analogous to that of Nagel and
 Schreckenberg. Recently, Chowdhury and Schadschneider (CS) incorporated the
 one-dimensional highway traffic model of Nagel and Schreckenberg as well as
 the two-dimensional BML model together to study microscopic dynamics of city
 traffic \cite{CS}.
\par
 It is not uncommon for local governments to set up tollbooths on highways. In
 fact, tolls can be used to control traffic flow and to increase revenue for
 local governments. Nevertheless, the presence of tollbooths will definitely
 slow down highway traffic. This is particularly true when tollbooths are set
 up on the highway rather than behind the entrance and exit ramps.
 Unfortunately, because of geographical and administrative considerations,
 tollbooths in many highways have to be built right on highways themselves. In
 fact, a few cellular automaton models of one-dimensional highway traffic flow
 with different kinds of blockages have been proposed and investigated
 \cite{One-D-Blockage}. They consider the effects of overtaking sites,
 bottleneck and quenched noise. In contrast, this paper investigates how the
 presence of tollbooths affects the traffic flow in a single lane highway using
 a simple-minded deterministic discrete model based on the cellular automaton
 traffic model of Fukui and Ishibashi \cite{FI} as well as the so-called green
 wave model of Torok and Kertesz \cite{Green_Wave}.
\par
 In their original model, Fukui and Ishibashi consider a one-dimensional array
 of $N$ sites with periodic boundary conditions. Each site may either be empty
 or have a single rightward moving car. They fix an integer $V_{\rm max}$ known
 as the maximum intrinsic car speed. Since their model does not consider the
 effect of car acceleration and deceleration, so at each timestep a car moves
 $k$ steps to the right where $k$ is the minimum of $V_{\rm max}$ and the
 number of consecutive empty sites to the left of the car. In addition, the
 motion of cars are updated in parallel \cite{FI}. They define the average car
 speed by
\begin{equation}
 \left< V \right> = \left< \frac{1}{N\rho} \sum_{i=1}^{N\rho} V_i \right> ~,
 \label{E:Def_V_Mean}
\end{equation}
 where $\rho$ is the car density in the system (and hence $N\rho$ is the total
 number of cars in the system) and $V_i$ is the speed of the $i$th car. Note
 that the right hand side of Eq.~(\ref{E:Def_V_Mean}) is averaged both over
 time and initial system configurations.
\par
 To coarse gain the highway system, we combine the FI model \cite{FI} with the
 so-called green wave traffic model \cite{Green_Wave}. That is to say, we call
 a collection of consecutive sites all containing rightward moving cars a car
 cluster. We demand that cars in the same cluster to move altogether as a group
 with speeds equal to that of the leading car in the group except possibly when
 the car passes through a tollbooth. Besides, all updates are taken in
 parallel. Finally, we introduce the effects of tollbooths by selecting
 $N_{\rm booth}$ special sites on the system. Whenever a car reaches these
 special tollbooth sites, it has to stop immediately and to wait for
 $t_{\rm wait}$ timesteps before it is allowed to move again. This waiting car,
 therefore, may block the motion of the cars queuing behind it for
 $t_{\rm wait}$ timesteps. For simplicity, tollbooths are located uniformly on
 the system.
\par
 For example, if $t_{\rm wait} = 0$, then a passing by car has to stop at the
 tollbooth site at once and then that car may move in the next timestep. Let
 us denote $0$ as a empty site, $1$ as a site occupied by a car, and an
 underline as a tollbooth site, then under our traffic rule,
 $1100101\underline{1}00$ will be transformed to $0011010\underline{1}01$ in
 the next timestep if $V_{\rm max} = 2$, $t_{\rm wait} = 0$.
\par
 Since we are interested only in the average car speed over all possible
 initial car configurations in the thermodynamic limit, therefore only the
 recurrent behavior of the system will affect the average car speed $\langle V
 \rangle$. More precisely, $\langle V\rangle$ depends only on the car density
 $\rho$, the maximum car speed $V_{\rm max}$, the tollbooth density
 $d_{\rm booth} \equiv N_{\rm booth} / N$ and the tollbooth stopping time
 $t_{\rm wait}$ of the system. Thus, $t_{\rm wait}$ and $d_{\rm booth}$ are the
 two controlling parameters in studying the effect of tollbooths.
\par
 In this paper, we are going to show that
\begin{Thrm}
 Case~(a): If $t_{\rm wait} = 0$, then the average car speed $\langle V\rangle$
 is given by
\begin{mathletters}
\begin{equation}
 \langle V\rangle = \left\{ \begin{array}{ll}
  \left( d_{\rm booth} \left\lceil \frac{1}{d_{\rm booth} V_{\rm max}}
   \right\rceil \right)^{-1} & \mbox{for~} 0 < \rho \leq \rho_a ,
   \vspace{1.5ex} \\
  1 / 2\rho & \mbox{for~} \rho_a < \rho \leq 1/2 , \vspace{1.5ex} \\
  1 & \mbox{for~} 1/2 < \rho < 1 , \vspace{1.5ex} \\
  0 & \mbox{for~} \rho = 1 ,
 \end{array} \right. \label{E:Vt0}
\end{equation}
 where $\rho_a = d_{\rm booth} \left\lceil 1 / d_{\rm booth} V_{\rm max}
 \right\rceil / 2$.
\par
 Case~(b): If $t_{\rm wait} > 0$, and $1/ d_{\rm booth} = 1 \bmod V_{\rm max}$
 then the average car speed is given by
\begin{equation}
 \langle V\rangle = \left\{ \begin{array}{ll}
  \left[ d_{\rm booth} \left( t_{\rm wait} + \left\lceil \frac{1}{d_{\rm booth}
   V_{\rm max}} \right\rceil \right) \right]^{-1} & \mbox{for~} 0 < \rho \leq
   \rho_b , \vspace{1.5ex} \\
  \left[ \rho \left( t_{\rm wait} + 1 \right) \right]^{-1} & \mbox{for~} \rho_b
   < \rho < 1 , \vspace{1.5ex} \\
  0 & \mbox{for~} \rho = 1 ,
 \end{array} \right. \label{E:VN1}
\end{equation}
 where $\rho_b = d_{\rm booth} ( t_{\rm wait} + \left\lceil 1 / d_{\rm booth}
 V_{\rm max} \right\rceil ) / (t_{\rm wait} + 1)$.
\par
 Case~(c): For the remaining possibility, namely that $t_{\rm wait} > 0$ and
 $1 / d_{\rm booth} \neq 1 \bmod V_{\rm max}$,, the average car speed is given
 by
\begin{equation}
 \langle V\rangle = \left\{ \begin{array}{ll}
  \left[ d_{\rm booth} \left( t_{\rm wait} + \left\lceil \frac{1}{d_{\rm booth}
   V_{\rm max}} \right\rceil \right) \right]^{-1} & \mbox{for~} 0 < \rho \leq
   \rho_{c1}, \vspace{1.5ex} \\ 
  \left[ \rho \left( t_{\rm wait} + 2 \right) \right]^{-1} & \mbox{for~}
   \rho_{c1} < \rho \leq \rho_{c2}, \vspace{1.5ex} \\
  \left[ d_{\rm booth} \left( t_{\rm wait} + 1 + \left\lceil
   \frac{1}{d_{\rm booth} V_{\rm max}} \right\rceil \right) \right]^{-1} &
   \mbox{for~} \rho_{c2} < \rho \leq \rho_{c3}, \vspace{1.5ex}
   \\
  \left[ \rho \left( t_{\rm wait} + 1 \right) \right]^{-1} & \mbox{for~}
   \rho_{c3} < \rho < 1, \vspace{1.5ex} \\
  0 & \mbox{for~} \rho = 1,
 \end{array} \right. \label{E:VNn}
\end{equation}
\end{mathletters}
 where $\rho_{c1} = d_{\rm booth} ( t_{\rm wait} + \left\lceil 1 /
 d_{\rm booth} V_{\rm max} \right\rceil ) / (t_{\rm wait} + 2)$, $\rho_{c2} =
 d_{\rm booth} ( t_{\rm wait} + 1 + \left\lceil 1 / d_{\rm booth} V_{\rm max}
 \right\rceil ) / (t_{\rm wait} + 2)$, $\rho_{c3} = d_{\rm booth} (t_{\rm wait}
 + 1 + \left\lceil 1 / d_{\rm booth} V_{\rm max} \right\rceil ) / (t_{\rm wait}
 + 1)$. \label{Thrm:V}
\end{Thrm}
\par
\begin{figure}
 \begin{center}
 \epsfxsize=5.8cm
 \begin{tabular}{lll}
 \epsfbox{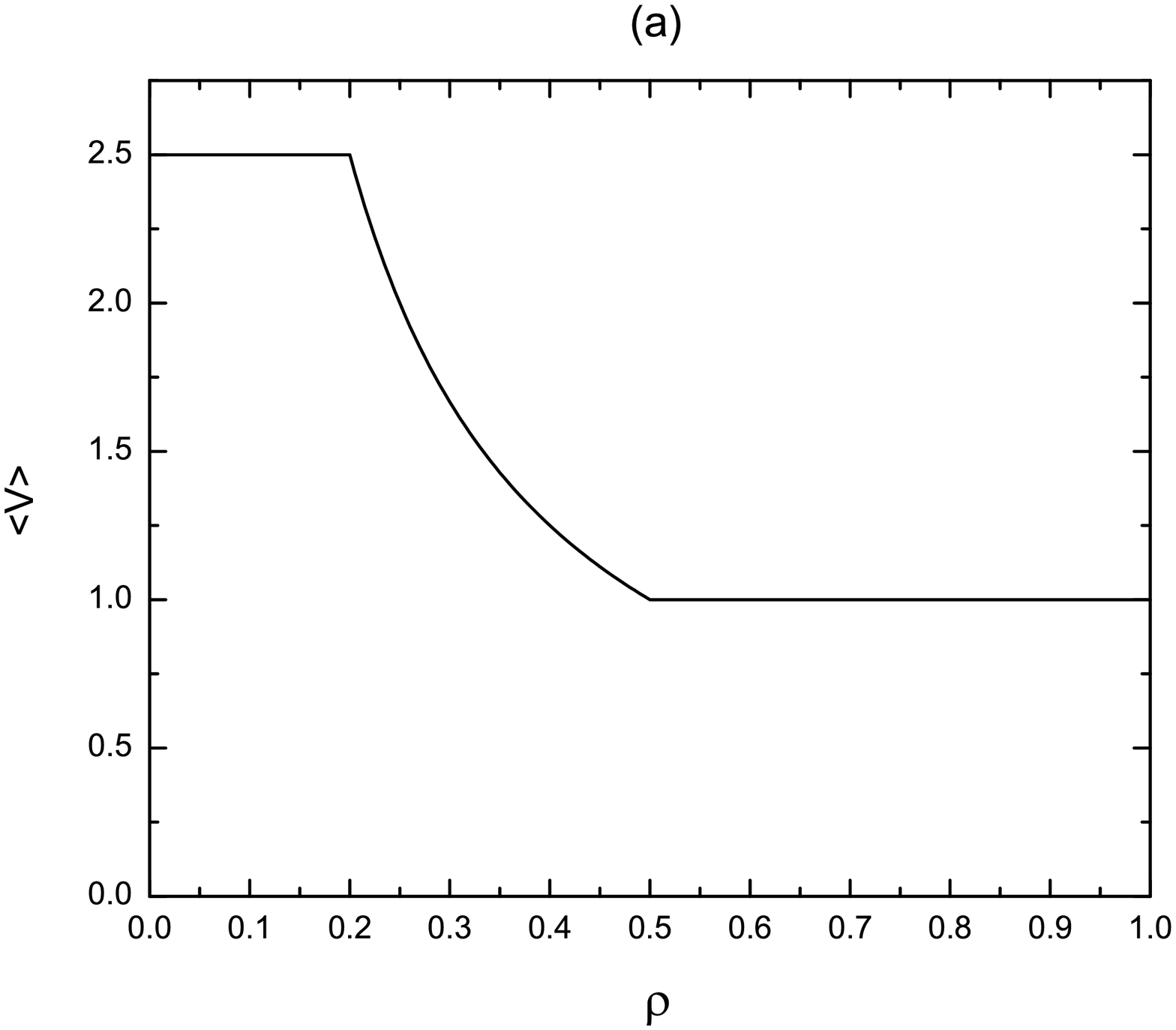} &
 \epsfbox{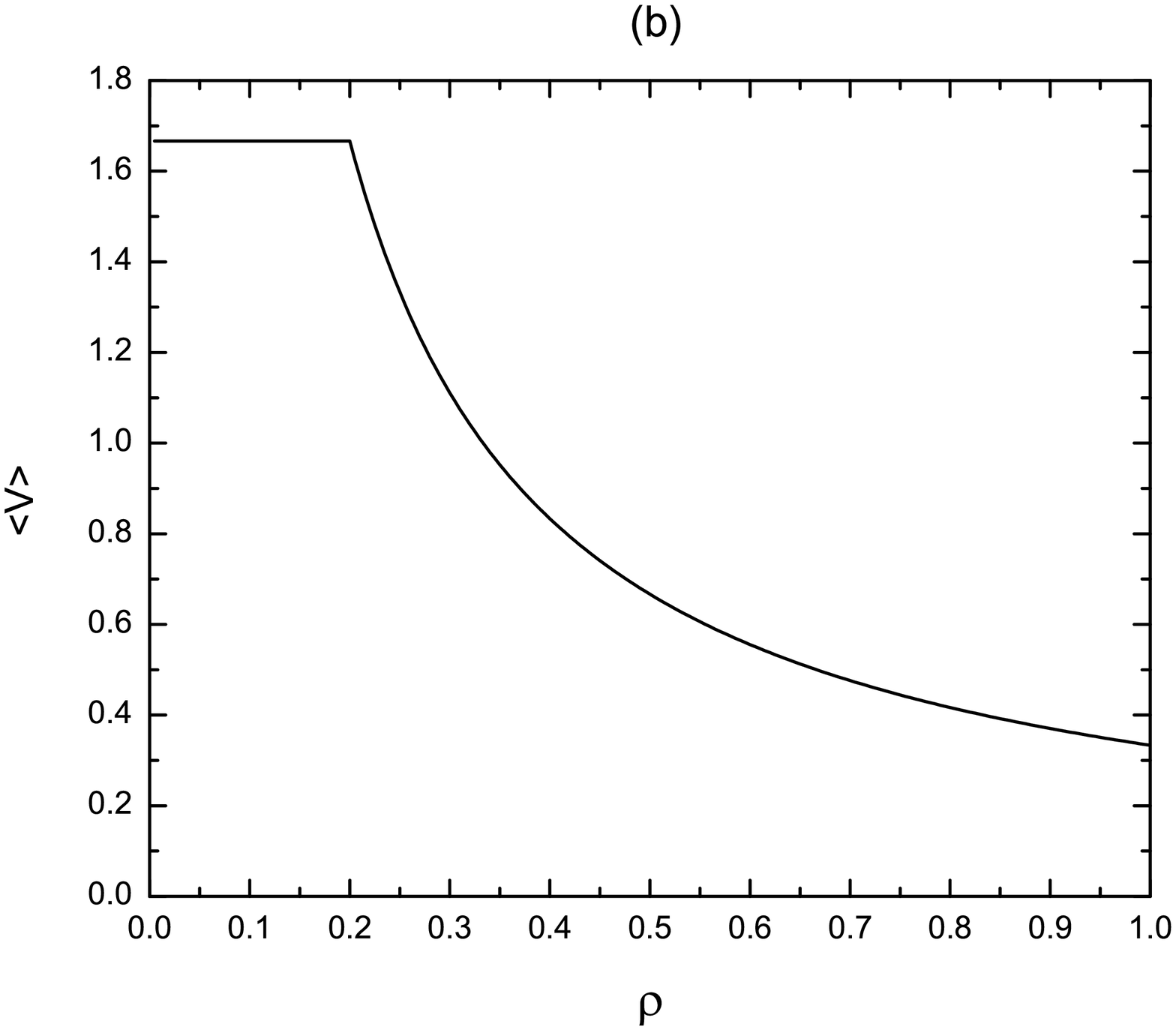} &
 \epsfbox{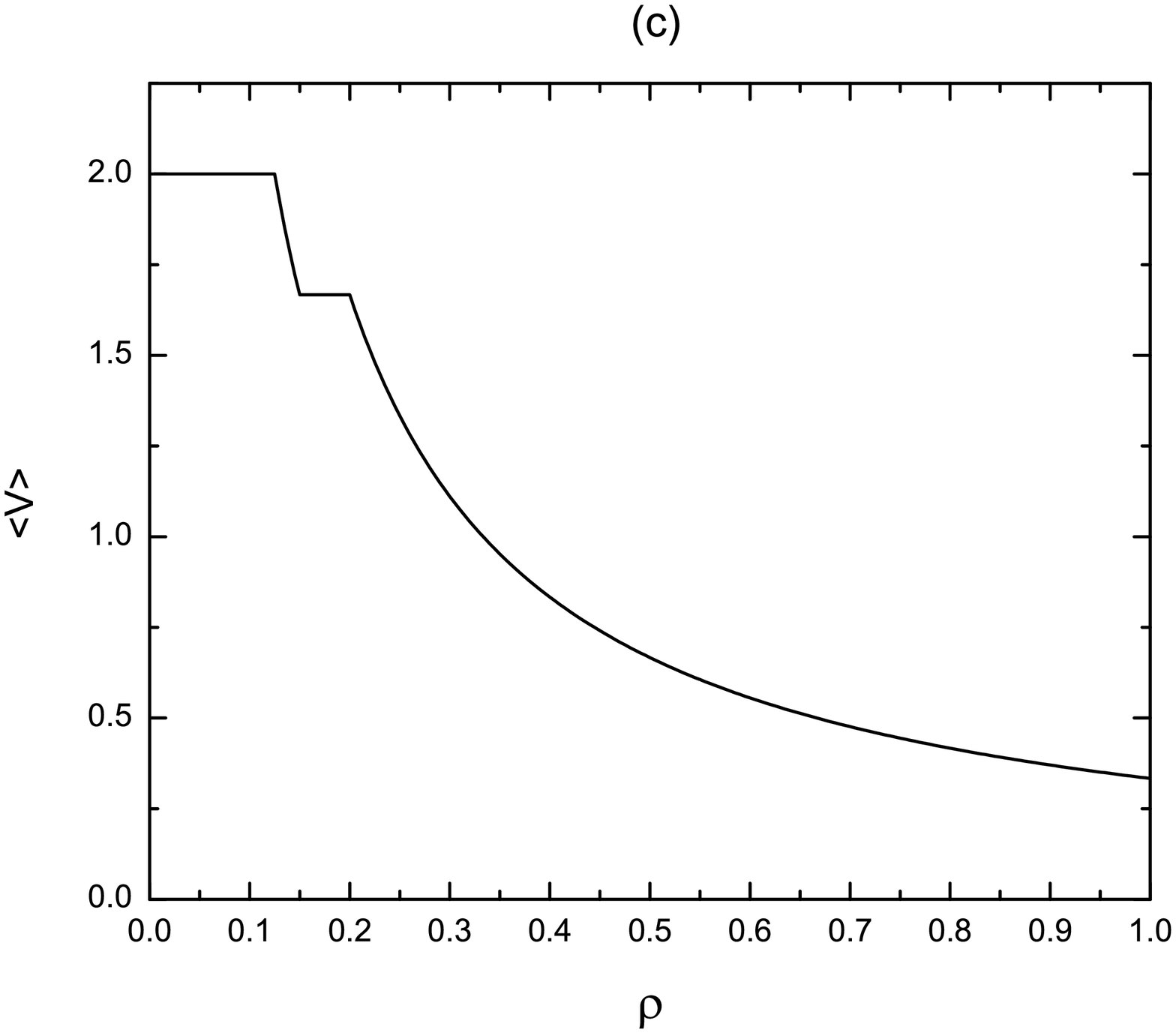}
 \end{tabular}
 \end{center}
 \caption{Typical shapes of the $\rho$ versus $\langle V\rangle$ curves in the
  three cases in Eqs.~(\ref{E:Vt0})--(\ref{E:VNn}). The parameters used in (a)
  (b) and (c) are ($t_{\rm wait} = 0$, $d_{\rm booth} = 1/10$, $V_{\rm max} =
  3$), ($t_{\rm wait} = 2$, $d_{\rm booth} = 1/10$, $V_{\rm max} = 3$) and
  ($t_{\rm wait} = 2$, $d_{\rm booth} = 1/10$, $V_{\rm max} = 4$) respectively.
 }
 \label{F:1}
\end{figure}

\medskip\indent
 Before going on to prove this theorem, we remark that in all the three cases
 above, first order phase transitions in $\langle V\rangle$ occur only at $\rho
 = 1$. All other transition points are second order in nature. (See
 Fig.~\ref{F:1}a--c for typical shapes of the $\rho$ versus $\langle V\rangle$
 curves in these three cases.)
\par\indent
 Now, we begin to prove this theorem by first considering the following Lemmas
 and Propositions.
\begin{Prop}
 The speed of cars in a recurrent state is never limited by another car
 provided that the car density of the system $\rho$ is less than or equals to
 $\rho_a$, $\rho_b$ and $\rho_{c1}$ for cases~(a), (b) and (c) in
 Theorem~\ref{Thrm:V} respectively. In other words, cars in systems with the
 above parameters move in maximum possible speed except possibly when they
 enter tollbooth sites. \label{Prop:HighSpeedPhase}
\end{Prop}
\par\noindent
Proof: According to our traffic rules, cars pass through a tollbooth site at a
 rate of at most one car every $(t_{\rm wait} + 1)$ timesteps. Moreover, a car
 cannot overtake one another and a car cluster can only break up at a tollbooth
 site. Thus, if there is a car in the recurrent state whose speed is limited by
 another car in front (towards the downstream direction), then all cars in the
 recurrent state must be blocked by the car in front therefore preventing them
 from moving in maximum speed $V_{\rm max}$ some time under the repeated
 application of our traffic rules. More importantly, the number of timesteps
 required to travel from one car to the next in a recurrent state under the
 action of our traffic rule cannot exceed $t_{\rm wait} + 1$. And, the blocking
 car must in turn being blocked in at least one of the previous $(t_{\rm wait}
 + 1)$ timesteps. Inductively, this can happen only when the average distance
 between cars in the configuration is greater than the minimum possible
 distance of a full speed recurrent configuration. But in a full speed
 recurrent configuration, a car can only be found at a tollbooth site, or at a
 site $k V_{\rm max}$ steps ahead of a tollbooth site where $k = 1,2,3,\ldots
 ,\left\lceil 1 / d_{\rm booth} V_{\rm max} \right\rceil$ and a car takes
 $(t_{\rm wait} + \left\lceil 1 / d_{\rm booth} V_{\rm max} \right\rceil )$
 timesteps to move through two consecutive tollbooths. Thus, there are
 $(t_{\rm wait} + \left\lceil 1 / d_{\rm booth} V_{\rm max} \right\rceil )$
 distinct possible ``locations'' for the cars in a recurrent state between two
 successive tollbooths if cars waiting at the same tollbooth for different
 times are regarded as distinct locations. And in order that cars do not block
 one another, an average of at most $0.5$, $\min ( 0.5,1/(t_{\rm wait} + 1)) =
 1/(t_{\rm wait} + 1)$ and $\min ( 0.5, 1/ (t_{\rm wait} + 2)) = 1 /
 (t_{\rm wait} + 2)$ of these ``locations'' are occupied in cases~(a), (b)
 and~(c) respectively. (Note that the difference between case~(b) and~(c) is
 due to the fact that in case~(b) an unobstructed car can move into an occupied
 tollbooth site at the same time when that occupying car moves out of that
 tollbooth site for $1 / d_{\rm booth} = 1 \bmod V_{\rm max}$. In contrast,
 this can never happen in case~(c).) Hence, in order that some car in the
 recurrent state is being blocked by another car, $\rho$ must be greater than
 $\rho_a$, $\rho_b$ and $\rho_{c1}$ in case~(a), (b) and~(c) respectively.
\hfill$\Box$
\begin{Lemma}
 Provided that $t_{\rm wait} > 0$ and $V_{\rm max} > 1$, a car in a recurrent
 state can only be blocked by a car cluster passing through a tollbooth.
 \label{Lemma:BlockPattern}
\end{Lemma}
\par\noindent
Proof: Suppose the contrary, we can find car A blocking car B in a recurrent
 state such that car A does not belong to any car cluster passing through a
 tollbooth. If we denote the action of applying our traffic rule once by
 ${\mathbf T}$, then the inverse map ${\mathbf T}^{-1}$ is well-defined on the
 set of all recurrent states. Clearly, under the action of ${\mathbf T}^{-1}$,
 car A must move backward by at least one site. Moreover, car A must be blocked
 by another car at least once in the previous $(t_{\rm wait} + 1)$ timesteps.
 Inductively, by considering the repeated application of ${\mathbf T}^{-1}$, we
 end up with a car C located at a tollbooth site whose motion in the next
 timestep is blocked by a car D located two sites in front. But this is
 impossible as the inverse image of this configuration under ${\mathbf T}^2$ is
 an empty set, contradicting the assumption that the state is recurrent.
\hfill$\Box$
\begin{Prop}
 In case~(c), whenever the car density lies between $\rho_{c1}$ and
 $\rho_{c2}$, a car in a recurrent state can only be blocked by another car
 right at a tollbooth site when the car occupying that tollbooth site is
 moving away. And whenever the car density lies between $\rho_{c2}$ and
 $\rho_{c3}$, a car in a recurrent state can only be blocked by another car
 right at a tollbooth site when the car occupying that tollbooth site is going
 to move away in the next timestep. \label{Prop:InterSpeedPhase}
\end{Prop}
\par\noindent
Proof: Proposition~\ref{Prop:HighSpeedPhase} implies that cars begin to block
 one another when $\rho > \rho_{c1}$ in case~(c). From
 Lemma~\ref{Lemma:BlockPattern} and the fact that $1/d_{\rm booth} \neq 1 \bmod
 V_{\rm max}$, we know that it is possible for a car A in a recurrent state to
 be blocked by a car located at a tollbooth site. If this event happens, car A
 has to take $(t_{\rm wait} + 1 + \left\lceil 1/d_{\rm booth} V_{\rm max}
 \right\rceil)$ timesteps to move through that two successive tollbooth sites.
 In other words, car A takes one timestep more than the minimum possible value
 in order to move through the two successive tollbooths. Note that if $\rho
 \leq \rho_{c3}$, no recurrent state can contain a car cluster C making up of
 more than two cars. The reason is simple: for otherwise, there exits, at any
 instance, at least one interval J between two tollbooths containing no more
 than $(t_{\rm wait} + \left\lceil 1 / d_{\rm wait} V_{\rm max} \right\rceil )/
 (t_{\rm wait} + 2)$ cars. But then the car outflow rate from this interval J
 is strictly less than one car per $(t_{\rm wait} + 1)$ timesteps while from
 Lemma~\ref{Lemma:BlockPattern} the car outflow rate from the cluster C equals
 one car every $(t_{\rm wait} + 1)$ timesteps. Thus eventually there is not
 enough car supply to maintain the car cluster C and new car clusters with more
 than two cars cannot be found elsewhere in the system due to the restrictions
 of both the can inflow and outflow rates in a tollbooth site. This contradicts
 our assumption that the configuration is recurrent.
\par
 Using the same trick as in the above argument that no car cluster with more
 than two cars can be formed in a recurrent configuration for $\rho \leq
 \rho_{c3}$, it is easy to show that when $\rho \in ( \rho_{c1}, \rho_{c2} ]$,
 the recurrent state consists of intervals of freely moving cars with density
 $\rho_{c1}$ as well as intervals of cars with density $\rho_{c2}$ that can be
 blocked by a ready-to-move car in a tollbooth site. (All car densities
 mentioned here are averaged over $(t_{\rm wait} + 2)$ timesteps.)
\par
 Similarly, when $\rho_{c2} < \rho \leq \rho_{c3}$, a recurrent state is made
 up of intervals of cars with density $\rho_{c2}$ that can be blocked by a
 ready-to-move car in a tollbooth site as well as intervals of cars with
 density $\rho_{c3}$ that can be blocked by a car in a tollbooth site that will
 move in the next timestep.
\hfill$\Box$
\begin{Lemma}
 Let $t_{\rm wait} = 0$ and $V_{\rm max} > 1$. Then if $\rho \leq 1/2$, all car
 clusters in a recurrent state consist of only one car. And if $\rho \geq 1/2$,
 the recurrent state contains no consecutive empty site.
 \label{Lemma:t0v1Pattern}
\end{Lemma}
\par\noindent
Proof: Since $t_{\rm wait} = 0$, all cars can move at least one step to the
 right in every timestep. Besides, it is not possible for two car clusters to
 merge. Since we are using green wave traffic rule and $V_{\rm max} > 1$, a
 car cluster may break up into two only at a tollbooth site. Furthermore,
 consecutive empty sites will ``move'' to the left while car clusters move to
 the right. Thus in $\mbox{O} (N)$ timesteps, the system will evolve to a state
 with maximum possible number of car clusters. Hence the lemma is proved.
\hfill$\Box$
\begin{Prop}
 Let $t_{\rm wait} = 0$, $V_{\rm max} > 1$ and $\rho_a \leq \rho < 1/2$, then
 exactly one car passes through a tollbooth every two timesteps.
 \label{Prop:t0v1Flow}
\end{Prop}
\par\noindent
Proof: From Lemma~\ref{Lemma:t0v1Pattern} and our traffic rule, we know that at
 most one car can pass through a tollbooth every two successive timesteps. And
 from Proposition~\ref{Prop:HighSpeedPhase}, we know that at least one car is
 blocked in a recurrent state. Using a similar argument as in the proof of
 Proposition~\ref{Prop:InterSpeedPhase}, one can always find a car whose motion
 is blocked by another about-to-go car locating at a tollbooth site in front.
 Hence, every car takes exactly two timesteps to pass through such a tollbooth.
\hfill$\Box$
\begin{Prop}
 The average car speed formulae in case~(b) and~(c) are valid for $\rho_b <
 \rho < 1$ and $\rho_{c3} < \rho < 1$, respectively. \label{Prop:LowSpeedPhase}
\end{Prop}
\par\noindent
Proof: From Lemma~\ref{Lemma:BlockPattern} and the assumption that $\rho_b <
 \rho < 1$ and $\rho_{c3} < \rho < 1$ in case~(b) and~(c) respectively, we can
 always find a car cluster in a recurrent state making up of at least three
 cars lining up in front of a tollbooth. Moreover, such a car cluster can never
 dissolve (that is, the cluster never disintegrates to clusters of single cars)
 under the repeated action of our traffic rule. Since the leading car in this
 car cluster passes through the tollbooth at a rate of once every
 $(t_{\rm wait} + 1)$ timesteps, we conclude that the average car speed
 $\langle V\rangle$ equals $1 / \rho (t_{\rm wait} + 1)$.
\hfill$\Box$ 
\par\medskip\noindent
Proof of Theorem~\ref{Thrm:V}: Since $\langle V\rangle$ is clearly equal to 0
 when $\rho = 1$, so it remains for us to consider the case when $\rho < 1$.
\par
 Let us first consider case~(a). If $V_{\rm max} = 1$, then our traffic rules
 reduce to moving each car forward one site at a time provided that $\rho < 1$.
 Hence, Eq.~(\ref{E:Vt0}) is trivially true in this case. So we only need to
 consider the case when $V_{\rm max} > 1$. And from
 Proposition~\ref{Prop:HighSpeedPhase}, we know that for case~(a),
 when $\rho \leq \rho_a$, cars will eventually moves as if there were no other
 cars in the system. Thus a car in a recurrent state takes $(t_{\rm wait} +
 \left\lceil 1 / d_{\rm booth} V_{\rm max} \right\rceil) = \left\lceil 1 /
 d_{\rm booth} V_{\rm max} \right\rceil$ timesteps to travel through
 $1/d_{\rm booth}$ sites. Hence, $\langle V\rangle = \left( d_{\rm booth}
 \left\lceil 1 / d_{\rm booth} V_{\rm max} \right\rceil \right)^{-1}$. And when
 $\rho_a < \rho \leq 1/2$, $\langle V\rangle = 1/2\rho$ is an immediate
 consequence of Proposition~\ref{Prop:t0v1Flow}. Finally, when $1/2 < \rho <
 1$, Lemma~\ref{Lemma:t0v1Pattern} tells us that two adjacent car clusters are
 separated by exactly one empty site. Hence, $\langle V\rangle = 1$ and
 Eq.~(\ref{E:Vt0}) holds for $t_{\rm wait} = 0$.
\par
 Now we are going prove Eq.~(\ref{E:VN1}) in case~(b). Clearly the validity of
 Eq.~(\ref{E:VN1}) for $\rho \leq \rho_b$ follows directly from
 Proposition~\ref{Prop:HighSpeedPhase}. And finally $\langle V\rangle$ for
 $\rho_b < \rho < 1$ has just been proven in
 Proposition~\ref{Prop:LowSpeedPhase}.
\par
 Lastly, the validity of Eq.~(\ref{E:VNn}) in case~(c) when $\rho \leq
 \rho_{c1}$ or $\rho > \rho_{c3}$ follows directly from
 Propositions~\ref{Prop:HighSpeedPhase} and~\ref{Prop:LowSpeedPhase}. And if
 $\rho_{c1} < \rho \leq \rho_{c2}$, Proposition~\ref{Prop:InterSpeedPhase}
 tells us that car density (averaged over $(t_{\rm wait} + 2)$ timesteps)
 between any two successive tollbooths in a recurrent state is either equal to
 $\rho_{c1}$ or $\rho_{c2}$. Thus, with a steady and continual supply of cars
 from behind, the car density averaged over $(t_{\rm wait} + 2)$ timesteps in
 every interval between two successive tollbooths is a constant. Since a car
 is released once every $(t_{\rm wait} + 2)$ timesteps from the tollbooth in a
 an interval with car density $\rho_{c2}$, hence $\langle V\rangle = 1/\rho
 (t_{\rm wait} + 2)$ in this car density range. Finally, when $\rho_{c2} < \rho
 < \rho_{c3}$, Proposition~\ref{Prop:InterSpeedPhase} implies that cars pass
 through every tollbooth once every $(t_{\rm wait} + 2)$ timesteps. Hence,
 Eq.~(\ref{E:VNn}) holds in this density range as well.
\hfill$\Box$
\par\medskip\indent
 In summary, we have investigated the behavior of a single lane deterministic
 highway traffic model in the presence of tollbooths. Our models are exactly
 solvable and the average car speed consists of a high speed, a partially
 jamming and a trivial completely jamming phases. The transition from the
 partially jamming phase to the completely jamming phase is first order in
 nature while all other transitions are second order.
\par
 Most importantly, our model suggests that the average car speed at high car
 density depends {\em only} on the car density and is independent of the detail
 arrangement of tollbooths in single lane traffic. While the regular placement
 of tollbooths and the deterministic traffic rules give rise to the
 unrealistically flat $\langle V\rangle$ when car density lies between
 $\rho_{c2}$ and $\rho_{c3}$, the general observation that for $t_{\rm wait} >
 0$, the average car speed $\langle V\rangle$ is approximately inversely
 proportional to $t_{\rm wait}$ is robust. Let us compare our results with the
 two-dimensional CS city traffic model. In the CS model, the time duration of
 red or green lights plays an analogous role of the tollbooths. As shown in
 their density versus flux per street curve in Fig~4 of Ref.~\cite{CS}, a
 linear region corresponding to cars moving with almost full speed is observed
 when the car density $\rho$ is low. More interestingly, a plateau region
 corresponding to the $\langle V\rangle \sim 1 / \rho$ is observed when the
 time duration of red or green lights is large and $\rho$ is around 0.1 to 0.4.
 And $\langle V\rangle$ starts to decrease at higher values of $\rho$ probably
 due to the effects of two-dimensional car blocking. Comparing the observations
 in the CS and our models, we believe that the $1/\rho$ behavior in high car
 density highway traffic is robust.
\acknowledgments
 H.F.C. and H.X. are supported in part by the Hong Kong SAR Government RGC
 Grant HKU~7098/00P and H.F.C. is also supported in part by the Outstanding
 Young Researcher Award of the University of Hong Kong.

\end{document}